\documentclass[12pt]{article}
\usepackage{epsf}
\title{ {\bf
$b$ quark Electric Dipole moment in the general two Higgs Doublet and 
three Higgs Doublet models}}
\author{\vspace{1cm}\\
        {\bf E. O. Iltan}
        \thanks{E-mail address:
        eiltan@heraklit.physics.metu.edu.tr}
 \\
        Physics Department, Middle East Technical University \\
        Ankara, Turkey\\}

\date{}

\begin{document}
\setlength{\baselineskip}{24pt}
\maketitle
\setlength{\baselineskip}{7mm}
\begin{abstract}
We study the Electric Dipole moment of $b$ quark in the general two Higgs
Doublet model (model III) and three Higgs Doublet model with $O(2)$ symmetry
in the Higgs sector. We  analyse the dependency of  this quantity  to the 
new phase coming from the complex Yukawa couplings and masses of charged 
and neutral Higgs bosons. We see that the Electric Dipole moment of $b$ 
quark is at the order of $10^{-20}\, e\,cm$, which is an extremely large 
value compared to one calculated in the SM and also two Higgs Doublet model 
(model II) with real Yukawa couplings.
\end{abstract} 
\thispagestyle{empty}
\newpage
\setcounter{page}{1}
\section{Introduction}
The study of CP violating effects provides comprehensive informations in the  
determination of free parameters of the various theoretical models. Non-zero 
Electric Dipole Moment (EDM) for elementary particles is the sign of such 
violation. Neutron EDM has a special interest and the experimental upper 
bound has been found as $d_N < 1.1\times 10^{-25} e\, cm$ \cite{smith}. 
EDM of electron and muon have been measured experimentally as 
$d_e =(-2.7\pm 8.3)\times 10^{-27} e\, cm$ \cite{abdullah} and  
$d_{\mu} =(3.7\pm 3.4)\times 10^{-19} e\, cm$ \cite{bailey}. These 
measurements can also give powerful clues about the internal structure 
of the particles, if it exists.  

The source of CP violation in the SM is complex Cabbibo-Kobayashi-
Maskawa (CKM) matrix elements. The quark EDM vanishes at one loop order
in the SM, since moduli of matrix element is involved in the relevant
expression. Further, it also vanishes at two loop order after sum over 
internal flavours \cite{sahab1,donog}. When QCD corrections are taken into 
account, nonzero EDM exists \cite{krause}. However, EDM for quarks is 
very small in the SM and need to be enhanced with the insertion of new 
models beyond. In the literature, quark EDM is calculated in the multi 
Higgs doublet models \cite{Weinberg, Branco, Smith}.  In \cite{Branco}, 
EDM due to the neutral Higgs boson effects is obtained in the two Higgs 
doublet model (2HDM) and \cite{Smith} discusses the necessity of more scalar 
fields than just two Higgs doublets for non-zero EDM when only the charged 
Higgs boson effects are taken into account. In \cite{Atwood1}, EDM of 
$b$-quark in the case of $Z$-boson and photon are calculated in the 2HDM and 
models with three and more Higgs doublets. EDM of b-quark for $Z$-boson is 
predicted in the range $10^{-21}-10^{-20} e\, cm$, following the scenario 
where CP violation may only come from the neutral Higgs sector. Further, 
$b$-quark EDM is obtained in the range $10^{-23}-10^{-22} e-cm$ when the CP 
violating effect comes from the charged sector. \cite{Xu} is devoted to the 
study of t-quark EDM, which is predicted at the order of the magnitude of 
$10^{-20} e\, cm$, arises from the neutral Higgs sector in the framework 
of the multi Higgs doublet model. In \cite{liao}, quark EDM is calculated 
in the 2HDM if the CP violating effects come from CKM matrix elements and 
it is thought that $H^{\pm}$ particles also mediate CP violation besides 
$W^{\pm}$ bosons, however, at the two loop order these new contributions 
vanish. The electric and weak electric dipole form factors for heavy
fermions in a general two Higgs doublet model is studied in \cite{Dumm} and 
it is concluded that the enhancement of three orders of magnitude in the 
electric dipole form factor of the $b$ quark with respect to the prediction 
of 2HDM I and II is possible 

In our work, we study EDM of $b$-quark in the general 2HDM (model III) and 
the general 3HDM with $O(2)$ symmetry in the Higgs sector ($3HDM(O_2)$)
\cite{eril3}. In this case, CP violation comes from complex Yukawa couplings 
and even at one loop, it is possible to get extremely large EDM, at the order 
of $10^{-20}\, e\, cm$. Since the theoretical values of quark EDM's are 
negligible in the SM and also in the model II 2HDM with real Yukawa
couplings, model III and $3HDM(O_2)$ results can give important information 
about the new parameters, namely masses of charged and neutral Higgs 
particles, Yukawa couplings, beyond the SM. Note that, non-zero EDM can 
be obtained due to the charged Higgs boson effects even in the 2HDM in 
this case.

The paper is organized as follows:
In Section 2, we present EDM for $b$ quark in the framework of model III and 
$3HDM(O_2)$. Section 3 is devoted to discussion and our conclusions.
\section{Electric Dipole moment of $b$ quark in the general two Higgs Doublet 
and three Higgs Doublet models} 
In this section, we calculate $b$-quark EDM in the model III and extend our 
results to the general 3HDM. Since  nonvanishing EDM is the sign for the CP
violation, we assume that there exist complex Yukawa couplings which are 
possible sources of such violations.  
We start with the Yukawa interaction in the model III
\begin{eqnarray}
{\cal{L}}_{Y}=\eta^{U}_{ij} \bar{Q}_{i L} \tilde{\phi_{1}} U_{j R}+
\eta^{D}_{ij} \bar{Q}_{i L} \phi_{1} D_{j R}+
\xi^{U\,\dagger}_{ij} \bar{Q}_{i L} \tilde{\phi_{2}} U_{j R}+
\xi^{D}_{ij} \bar{Q}_{i L} \phi_{2} D_{j R} + h.c. \,\,\, ,
\label{lagrangian}
\end{eqnarray}
where $L$ and $R$ denote chiral projections $L(R)=1/2(1\mp \gamma_5)$,
$\phi_{i}$ for $i=1,2$, are the two scalar doublets, 
$\bar{Q}_{i L}$ are left handed quark doublets, $U_{j R} (D_{j R})$ are 
right handed up (down) quark singlets, with  family indices $i,j$. The 
Yukawa matrices $\eta^{U,D}_{ij}$ and $\xi^{U,D}_{ij}$ have in general 
complex entries. Here $\phi_{1}$ and $\phi_{2}$ are chosen as
\begin{eqnarray}
\phi_{1}=\frac{1}{\sqrt{2}}\left[\left(\begin{array}{c c} 
0\\v+H^{0}\end{array}\right)\; + \left(\begin{array}{c c} 
\sqrt{2} \chi^{+}\\ i \chi^{0}\end{array}\right) \right]\, ; 
\phi_{2}=\frac{1}{\sqrt{2}}\left(\begin{array}{c c} 
\sqrt{2} H^{+}\\ H_1+i H_2 \end{array}\right) \,\, .
\label{choice}
\end{eqnarray}
with the vacuum expectation values,  
\begin{eqnarray}
<\phi_{1}>=\frac{1}{\sqrt{2}}\left(\begin{array}{c c} 
0\\v\end{array}\right) \,  \, ; 
<\phi_{2}>=0 \,\, ,
\label{choice2}
\end{eqnarray}
to collect SM particles in the first doublet and new particles in the second 
one. The Flavor Changing (FC) part of the interaction can be written as 
\begin{eqnarray}
{\cal{L}}_{Y,FC}=
\xi^{U\,\dagger}_{ij} \bar{Q}_{i L} \tilde{\phi_{2}} U_{j R}+
\xi^{D}_{ij} \bar{Q}_{i L} \phi_{2} D_{j R} + h.c. \,\, ,
\label{lagrangianFC}
\end{eqnarray}
where the couplings  $\xi^{U,D}$ for the FC charged interactions are
\begin{eqnarray}
\xi^{U}_{ch}&=& \xi^U_{N} \,\, V_{CKM} \nonumber \,\, ,\\
\xi^{D}_{ch}&=& V_{CKM} \,\, \xi^D_{N} \,\, ,
\label{ksi1} 
\end{eqnarray}
and $\xi^{U,D}_{N}$ is defined by the expression (for more details see 
\cite{soni})
\begin{eqnarray}
\xi^{U (D)}_{N}=(V_{R (L)}^{U (D)})^{-1} \xi^{U,(D)} V_{L(R)}^{U (D)}\,\, .
\label{ksineut}
\end{eqnarray}
Note that the index "N" in $\xi^{U,D}_{N}$ denotes the word "neutral". 

\begin{figure}[htb]
\vskip -3.0truein
\centering
\epsfxsize=6.8in
\leavevmode\epsffile{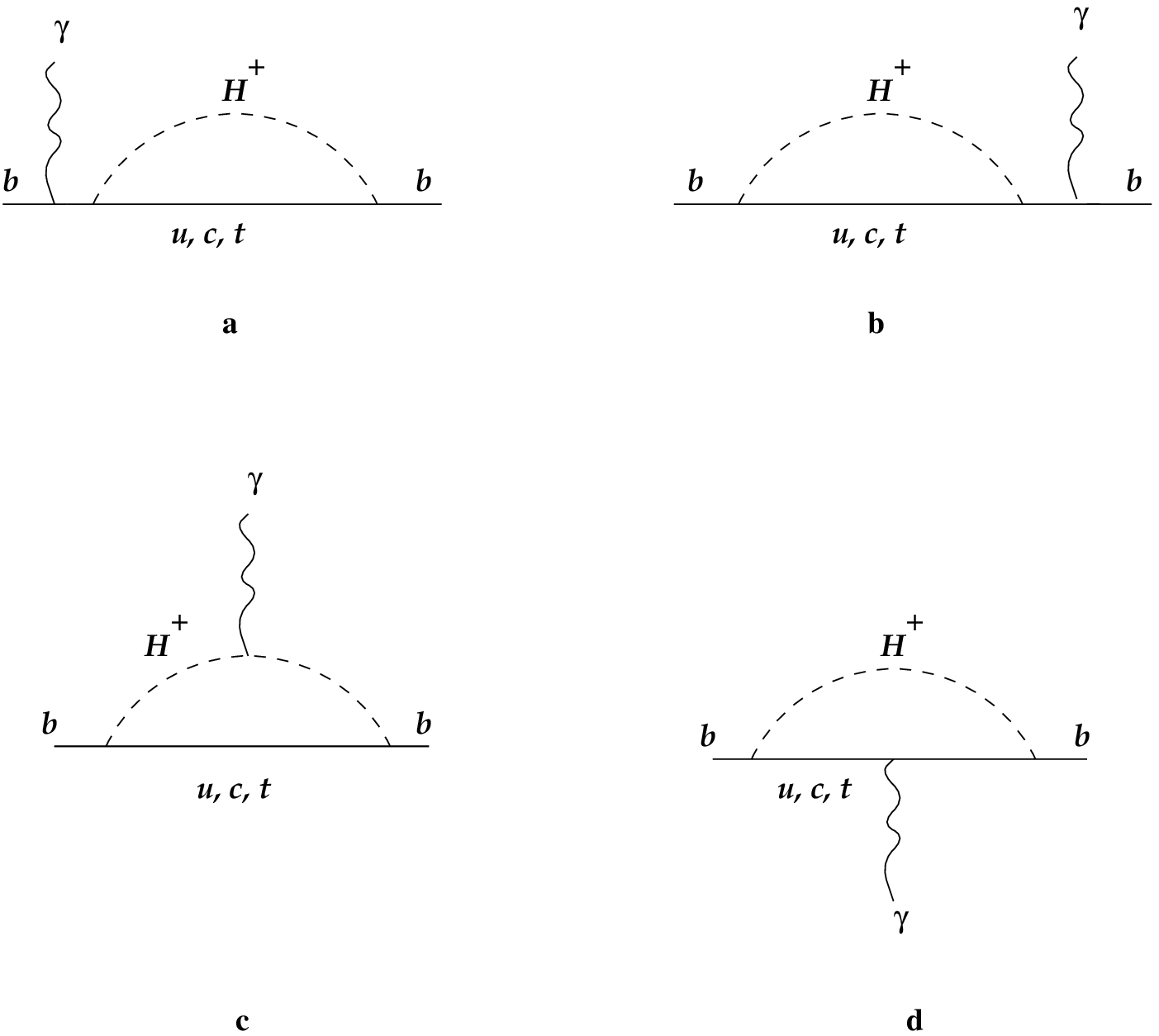}
\vskip -7.0truein
\caption[]{One loop diagrams contribute to EDM of $b$-quark due to 
$H^{\pm}$ in the 2HDM. Wavy lines represent the electromagnetic field and 
dashed lines the $H^{\pm}$ field.}
\label{fig1}
\end{figure}

\begin{figure}[htb]
\vskip -3.0truein
\centering
\epsfxsize=6.8in
\leavevmode\epsffile{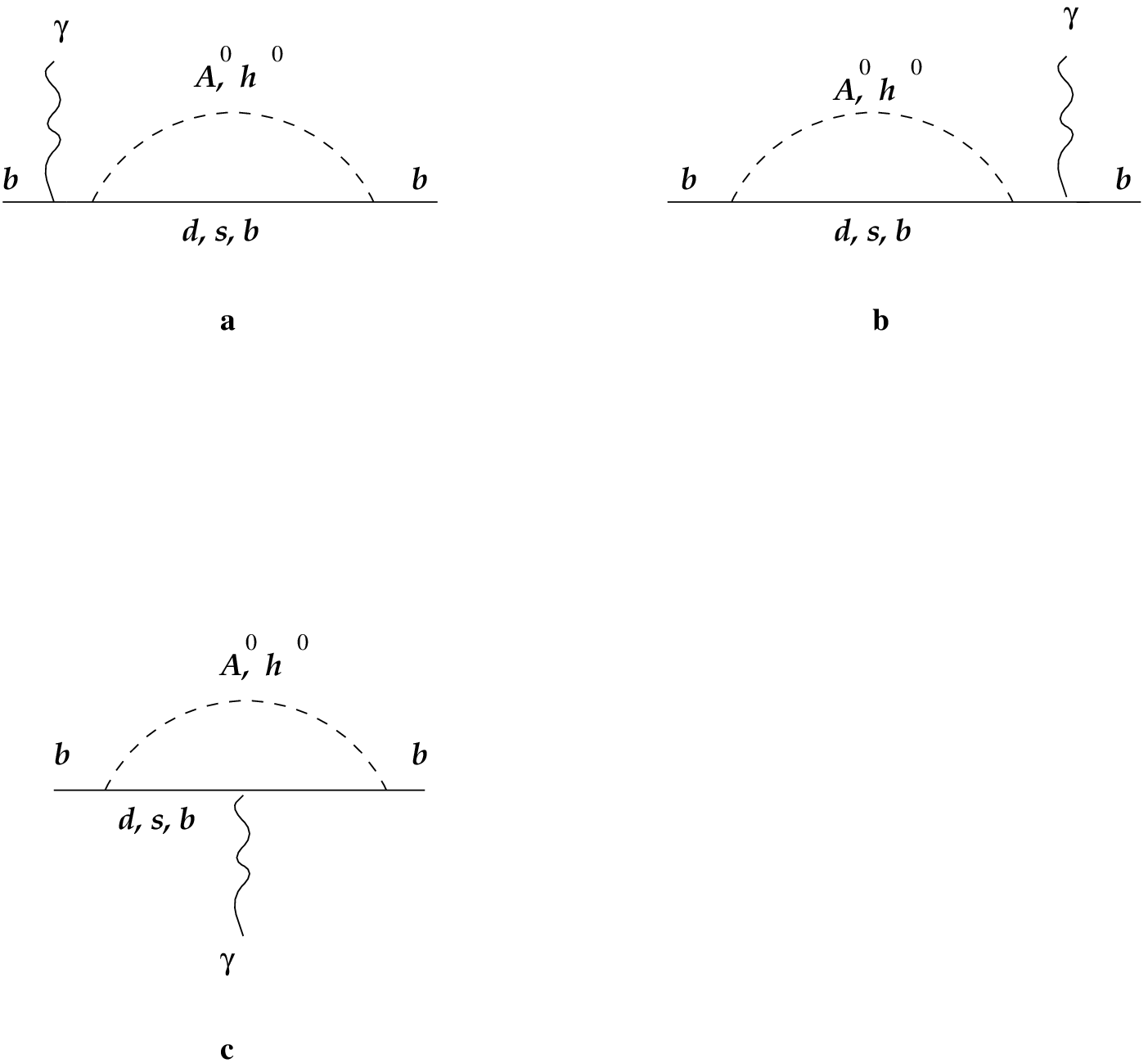}
\vskip -7.0truein
\caption[]{The same as Fig. \ref{fig1}, but for neutral Higgs bosons 
$h^0$ and $A^0$.}
\label{fig2}
\end{figure}

The effective EDM interaction for $b$-quark is defined as 
\begin{eqnarray}
{\cal L}_{EDM}=i d \,\bar{b}\,\gamma_5 \,\sigma_{\mu\nu}\,b\, F^{\mu\nu} \,\, ,
\label{EDM1}  
\end{eqnarray}
where $F_{\mu\nu}$ is the electromagnetic field tensor and "$d$" is EDM of 
$b$-quark. Here, "$d$" is a real number by hermiticity. In model 
III, the charged $H^{\pm}$ and neutral Higgs bosons $h^0, A^0$ can induce 
CP violating interactions which can create EDM at loop level. Note that, we
take $H_1$ and $H_2$ as the mass eigenstates $h^0$ and $A^0$ respectively 
and do not take the mixing of two CP-even neutral bosons $H_0$ and $h_0$ 
into account, since no mixing occurs at tree level for our choice of Higgs 
doublets (eq. (\ref{choice})). We present the necessary 1-loop diagrams due 
to charged and neutral Higgs particles in Figs. \ref{fig1} and \ref{fig2}. 
Since, in the on-shell renormalization scheme, the self energy $\sum(p)$ 
can be written as 
\begin{eqnarray}
\sum(p)=(\hat{p}-m_b)\bar{\sum}(p) (\hat{p}-m_b)\,\, ,
\label{self}
\end{eqnarray}
diagrams $a$, $b$ in Figs. \ref{fig1} and \ref{fig2} vanish when 
$b$-quark is on-shell. However the vertex diagrams $c$, $d$ in Fig. 
\ref{fig1} and $c$ in Fig. \ref{fig2} give non-zero contributions. 
The most general vertex operator for on-shell $b$-quark can be written as
\begin{eqnarray}
\Gamma_{\mu}&=&F_1(q^2)\, \gamma_{\mu}+ F_2 (q^2)\, \sigma_{\mu\nu} \,
q^{\nu}\nonumber \\ 
&+& F_3 (q^2)\, \sigma_{\mu\nu}\gamma_5\, q^{\nu}
\label{vertexop}
\end{eqnarray}
where $q_{\nu}$ is photon 4-vector and $q^2$ dependent form factors 
$F_{1}(q^2)$ and  $F_{2}(q^2)$ are proportional to the charge and 
anamolous magnetic moment of $b$-quark respectively. CP
violated interaction exists with the non-zero value of $F_{3}(q^2)$ and 
it is proportional to EDM of $b$-quark. By extracting the CP violating
part of the vertex, $b$-quark EDM "$d$" (see eq.(\ref{EDM1})) is calculated 
as a sum of contributions coming from charged and neutral Higgs bosons,
\begin{eqnarray}
d=d^{H^{\pm}}+d^{h^0}+d^{A^0}\,\,,
\label{EDM2}
\end{eqnarray}
where $d^{H^{\pm}}$ , $d^{h^0}$ and $d^{A^0}$ are
\begin{eqnarray}
d^{H^{\pm}}&=&\frac{4\,G_F}{\sqrt {2}}
\frac{e}{32\pi^2}\, \frac{1}{m_t}\, \bar{\xi}^{U}_{N,tt}\,
Im(\bar{\xi}^{D}_{N,bb})\, |V_{tb}|^2 \, 
\frac{y\, \big( (-1\,+\,Q_t\, (-3\,+\,y)-\,y)(y\,-\,1)+2\, (Q_t\,+\,y) 
\,ln\,y \big)} {(y-1)^3}
\,\, , \nonumber \\
d^{h^0}&=&-\frac{4\,G_F}{\sqrt {2}}
\frac{e}{16\pi^2}\, \frac{Q_b}{m_b}\, Im(\bar{\xi}^{D}_{N,bb})
\,  Re (\bar{\xi}^{D}_{N,bb})\, 
(1-\frac{r_1\,(r_1-2)}{\sqrt{r_1\,(r_1-4)}}\,
ln\frac{\sqrt{r_1}-\sqrt{r_1-4}}{2}-\frac{1}{2} r_1 \,ln\, r_1 )
\,\, , \nonumber \\
d^{A^0}&=&-d^{h^0} (r_1\rightarrow r_2)
\,\, ,
\label{EDM3}
\end{eqnarray}
with $r_1=m_{h^0}^2/m_b^2$, $r_2=m_{A^0}^2/m_b^2$ and 
$y=\frac{m_t^2}{m^2_{H^{\pm}}}$. Here $Q_b$ and $Q_t$ are charges of 
$b$ and $t$ quarks respectively and 
$\xi^{U(D)}_{N,ij}=\sqrt{\frac{4\, G_F}{\sqrt{2}}}\,
\bar{\xi}^{U(D)}_{N,ij}$. In eq. (\ref{EDM3}) we take into account 
only internal $t$-quark contribution for charged Higgs and internal 
$b$-quark contribution for neutral Higgs interactions since we assume that 
the Yukawa couplings $\bar{\xi}^{U}_{N,ib},\, i=u,c $,
$\bar{\xi}^{D}_{N, bj},\, j=d,s $ and  $\bar{\xi}^{U}_{N, tc}$ are 
negligible compared to $\bar{\xi}^{U}_{N,tt}$ and $\bar{\xi}^{D}_{N,bb}$ 
(see \cite{eril2}). Further, we choose $\bar{\xi}^{U}_{N,tt}$ real and 
$\bar{\xi}^{D}_{N,bb}$ complex, 
\begin{eqnarray}
\bar{\xi}^{D}_{N,bb}=|\bar{\xi}^{D}_{N,bb}|\, e^{i \theta}\, . 
\label{xi}
\end{eqnarray}

Finally, using the parametrization eq.(\ref{xi}), we get the EDM of $b$-quark 
in model III as 
\begin{eqnarray}
d&=&\frac{4\,G_F}{\sqrt {2}} \frac{e}{32\pi^2}\, 
|\bar{\xi}^{D}_{N,bb}|\, sin\,\theta \{ 
\frac{1}{m_t}\, \bar{\xi}^{U}_{N,tt}\, |V_{tb}|^2 \, 
\frac{y\, \big( (-1\,+\,Q_t\, (-3\,+\,y)-\,y)(y\,-\,1)+2\, 
(Q_t\,+\,y) \,ln\,y\big)}{(y-1)^3} 
\nonumber \\ &+& 
2 \frac{Q_b}{m_b}\, |\bar{\xi}^{D}_{N,bb}| \,cos\,\theta 
\big(
\frac{r_1\,(r_1-2)\,}{\sqrt{r_1\,(r_1-4)}}\, 
ln\frac{\sqrt{r_1}-\sqrt{r_1-4}}{2}+\frac{1}{2} r_1\, ln\, r_1 
\nonumber \\ &-&
\frac{r_2\,(r_2-2)\,}{\sqrt{r_2\,(r_2-4)}}\, 
ln\frac{\sqrt{r_2}-\sqrt{r_2-4}}{2}-\frac{1}{2} r_2\, ln\, r_2 
\big )  \}\,\, .
\label{EDM4}
\end{eqnarray}

Now, we would like to extend our result to the general $3HDM(O_2)$ 
\cite{eril3}. The new general Yukawa interaction is 
\begin{eqnarray}
{\cal{L}}_{Y}&=&\eta^{U}_{ij} \bar{Q}_{i L} \tilde{\phi_{1}} U_{j R}+
\eta^{D}_{ij} \bar{Q}_{i L} \phi_{1} D_{j R}+
\xi^{U\,\dagger}_{ij} \bar{Q}_{i L} \tilde{\phi_{2}} U_{j R}+
\xi^{D}_{ij} \bar{Q}_{i L} \phi_{2} D_{j R} \nonumber \\
&+&
\rho^{U\,\dagger}_{ij} \bar{Q}_{i L} \tilde{\phi_{3}} U_{j R}+
\rho^{D}_{ij} \bar{Q}_{i L} \phi_{3} D_{j R}
 + h.c. \,\,\, ,
\label{lagrangian3H}
\end{eqnarray}
where $\phi_{i}$ for $i=1,2,3$, are three scalar doublets and  
$\eta^{U,D}_{ij}$, $\xi^{U,D}_{ij}$, $\rho^{U,D}_{ij}$ are
the Yukawa matrices having complex entries, in general. 
With the choice 
\begin{eqnarray}
\phi_{1}=\frac{1}{\sqrt{2}}\left[\left(\begin{array}{c c} 
0\\v+H^{0}\end{array}\right)\; + \left(\begin{array}{c c} 
\sqrt{2} \chi^{+}\\ i \chi^{0}\end{array}\right) \right]\, , 
\nonumber \\ \\
\phi_{2}=\frac{1}{\sqrt{2}}\left(\begin{array}{c c} 
\sqrt{2} H^{+}\\ H^1+i H^2 \end{array}\right) \,\, ,\,\, 
\phi_{3}=\frac{1}{\sqrt{2}}\left(\begin{array}{c c} 
\sqrt{2} F^{+}\\ H^3+i H^4 \end{array}\right) \,\, ,\nonumber
\label{choice3H}
\end{eqnarray}
and the vacuum expectation values,  
\begin{eqnarray}
<\phi_{1}>=\frac{1}{\sqrt{2}}\left(\begin{array}{c c} 
0\\v\end{array}\right) \,  \, ; 
<\phi_{2}>=0 \,\, ; <\phi_{3}>=0\,\,  
\label{choice23H}
\end{eqnarray}
the information about new physics beyond the SM is carried by the last 
two doublets $\phi_2$ and $\phi_3$, while the first doublet $\phi_1$ 
describes only the SM part . The Yukawa interaction for the Flavor 
Changing (FC) part is
\begin{eqnarray}
{\cal{L}}_{Y,FC}=
\xi^{U\,\dagger}_{ij} \bar{Q}_{i L} \tilde{\phi_{2}} U_{j R}+
\xi^{D}_{ij} \bar{Q}_{i L} \phi_{2} D_{j R}
+\rho^{U\,\dagger}_{ij} \bar{Q}_{i L} \tilde{\phi_{3}} U_{j R}+
\rho^{D}_{ij} \bar{Q}_{i L} \phi_{3} D_{j R} + h.c. \,\, ,
\label{lagrangianFC3H}
\end{eqnarray}
where, the charged couplings  $\xi_{ch}^{U,D}$ and $\rho_{ch}^{U,D}$ are 
\begin{eqnarray}
\xi^{U}_{ch}&=& \xi_{N} \,\, V_{CKM} \nonumber \,\, ,\\
\xi^{D}_{ch}&=& V_{CKM} \,\, \xi_{N}  \nonumber \,\, , \\
\rho^{U}_{ch}&=& \rho_{N} \,\, V_{CKM} \nonumber \,\, ,\\
\rho^{D}_{ch}&=& V_{CKM} \,\, \rho_{N} \,\, ,
\label{ksi13H} 
\end{eqnarray}
and
\begin{eqnarray}
\xi^{U (D)}_{N}=(V_{R(L)}^{U (D)})^{-1} \xi^{U,(D)} V_{L(R)}^{U (D)}\,\, , 
\nonumber \\
\rho^{U (D)}_{N}=(V_{R(L)}^{U (D)})^{-1} \rho^{U,(D)} V_{L(R)}^{U (D)}
\,\, .
\label{ksineut3H}
\end{eqnarray}
In the 3HDM, there are additional charged Higgs particles ,$F^{\pm}$,  and
neutral Higgs bosons $h^{\prime\, 0}$, $A^{\prime\, 0}$ (see \cite {eril3}). 
These particles bring new part to EDM of $b$-quark and by taking only 
couplings $\bar{\xi}^U_{N,tt}$ ,$\bar{\xi}^D_{N,bb}$, 
$\bar{\rho}^U_{N,tt}$ and  $\bar{\rho}^D_{N,bb}$ into account, it can be 
written as 
\begin{eqnarray}
d=d^{H^{\pm}}+d^{h^0}+d^{A^0}+d^{F^{\pm}}+d^{h^{\prime 0}}+
d^{A^{\prime 0}}\,\,,
\label{EDM23H}
\end{eqnarray}
where $d^{F^{\pm}}$ , $d^{h^{\prime 0}}$ and $d^{A^{\prime 0}}$ are
\begin{eqnarray}
d^{F^{\pm}}&=&\frac{4\,G_F}{\sqrt {2}} 
\frac{e}{32\pi^2}\, \frac{1}{m_t}\, \bar{\rho}^{U}_{N,tt}\,
Im(\bar{\rho}^{D}_{N,bb})\, |V_{tb}|^2 \, 
\frac{y\, ( (-1\,+\,Q_t\, (-3\,+\,y')-\,y')(y'\,-\,1)+2\, (Q_t\,+\,y') 
\,ln\,y')}{(y'-1)^3}
\,\, , \nonumber \\
d^{h^{\prime 0}}&=&-\frac{4\,G_F}{\sqrt {2}} 
\frac{e}{16\pi^2}\, \frac{Q_b}{m_b}\, 
Im(\bar{\rho}^{D}_{N,bb})\,  Re (\bar{\rho}^{D}_{N,bb})\, 
(1-\frac{r^{\prime}_1\,(r^{\prime}_1-2)\,}
{\sqrt{r^{\prime}_1\,(r^{\prime}_1-4)}}\, 
ln\frac{\sqrt{r^{\prime}_1}-\sqrt{r^{\prime}_1-4}}{2}-
\frac{1}{2} r^{\prime}_1\, ln\, r^{\prime}_1 )
\,\, , \nonumber \\
d^{A^{\prime 0}}&=&-d^{h^0} (r^{\prime}_1\rightarrow r^{\prime}_2)
\,\, ,
\label{EDM3H1}
\end{eqnarray}
with $r^{\prime}_1=m_{h^{\prime 0}}^2/m_b^2$, $r_2=m_{A^{\prime 0}}^2/m_b^2$ 
and $y'=\frac{m_t^2}{m_{F^{\pm}}^2}$. The expressions for 
$d^{H^{\pm}}$, $d^{h^{0}}$ and $d^{A^{0}}$ are defined in eq.(\ref{EDM3}). 
Further, by introducing $O(2)$ symmetry in the Higgs sector,
the masses of new Higgs bosons $F^{\pm}$, $h^{\prime \,0}$ and 
$A^{\prime \,0}$ become  the same as the masses of model III Higgs bosons, 
$H^{\pm}$, $h^{0}$ and  $A^{0}$ respectively (see \cite {eril3}).
Therefore, the final result for EDM of $b$-quark in the $3HDM(O_2)$ is 
\begin{eqnarray}
d&=&\frac{4\,G_F}{\sqrt {2}} \frac{e}{32\pi^2}
\{
\frac{1}{m_t}\, 
(\bar{\xi}^{U}_{N,tt}\,Im(\bar{\xi}^{D}_{N,bb})\, + 
\bar{\rho}^{U}_{N,tt}\,Im(\bar{\rho}^{D}_{N,bb}) )\,|V_{tb}|^2 
\nonumber \\ & &
\frac{y\, ( (-1\,+\,Q_t\, (-3\,+\,y)-\,y)(y\,-\,1)+2\, (Q_t\,+\,y) \,ln\,y)}
{(y-1)^3}
\,\,  \nonumber \\&+& 
2 \, \frac{Q_b}{m_b}\, 
(Im(\bar{\xi}^{D}_{N,bb})\,  Re (\bar{\xi}^{D}_{N,bb})\,+
Im(\bar{\rho}^{D}_{N,bb})\,  Re (\bar{\rho}^{D}_{N,bb})) 
\nonumber \\ & &
(\frac{r_1\,(r_1-2)}{\sqrt{r_1\,(r_1-4)}}\,
ln\frac{\sqrt{r_1}-\sqrt{r_1-4}}{2}+\frac{1}{2} r_1\, ln\, r_1 
\nonumber \\ &-& 
\frac{r_2\,(r_2-2)}{\sqrt{r_2\,(r_2-4)}}\,
ln\frac{\sqrt{r_2}-\sqrt{r_2-4}}{2}-\frac{1}{2} r_2\, ln\, r_2) \}\, .
\label{EDM3H2}
\end{eqnarray}
New $O(2)$ symmetry also permits us to parametrize the Yukawa matrices 
$\bar{\xi}^{U(D)}$ and $\bar{\rho}^{U(D)}$ as  \cite {eril3}
\begin{eqnarray}
\bar{\xi}^{U (D)}=\epsilon^{U(D)} cos\,\theta \nonumber \,\, ,\\
\bar{\rho}^{U}=\epsilon^{U} sin\,\theta \nonumber \,\, ,\\ 
\bar{\rho}^{D}=i \epsilon^{D} sin\,\theta \,\, ,
\label{yukpar}
\end{eqnarray}
where $\epsilon^{U(D)}$ are real matrices satisfy the equation 
\begin{eqnarray}
(\bar{\xi}^{\prime U(D)})^+ \bar{\xi}^{\prime U (D) } +
(\bar{\rho}^{\prime U (D)})^+\bar{\rho}^{\prime U (D) }=
(\epsilon^{U(D)})^T \epsilon^{U(D)} 
\label{yukpareq}
\end{eqnarray}
Here $T$ denotes transpose operation. In eq. (\ref{yukpar}),  we take 
$\bar{\rho}^{D}$ complex to carry all $CP$ violating effects on the third 
Higgs scalar.  Using the parametrization in eq. (\ref{yukpar}), it is easy to 
see that only charged Higgs bosons contribute to EDM of $b$-quark but not the
neutral Higgs ones, since $\bar{\rho}^{D}_{N,bb}$ is a pure imaginary
number. Finally, $b$-quark EDM reads as
\begin{eqnarray}
d=\frac{4\,G_F}{\sqrt {2}} \frac{e}{32\pi^2} \frac{1}{m_t}\,|V_{tb}|^2 \, 
\bar{\epsilon}^{U}_{N,tt}\,\bar{\epsilon}^{D}_{N,bb}\, sin^2\,\theta 
\frac{y\, ( (-1\,+\,Q_t\, (-3\,+\,y)-\,y)(y\,-\,1)+2\, (Q_t\,+\,y) \,ln\,y)}
{(y-1)^3}
\label{EDM3H3}
\end{eqnarray}
%
\section{Discussion}
In this section, we study dependencies of $b$-quark EDM on the masses of 
charged and neutral Higgs bosons and the CP violating parameter $\theta$.
In the analysis, we use the CLEO measurement \cite{cleo2} 
\begin{eqnarray}
Br (B\rightarrow X_s\gamma)= (3.15\pm 0.35\pm 0.32)\, 10^{-4} \,\, .
\label{br2}
\end{eqnarray}
to find a constraint region for our free parameters. The procedure is to 
restrict the Wilson coefficient $C_7^{eff}$, which is the effective 
coefficient of the operator $O_7 = \frac{e}{16 \pi^2} \bar{s}_{\alpha} 
\sigma_{\mu \nu} (m_b R + m_s L) b_{\alpha} {\cal{F}}^{\mu \nu}$
(see \cite{alil1} and references therein), in the region 
$0.257 \leq |C_7^{eff}| \leq 0.439$. Here upper and lower limits were 
calculated using the CLEO measurement and all possible uncertainities in the 
calculation of $C_7^{eff}$ \cite{alil1}. This restriction allows us to find 
a region for the parameters $\bar{\xi}^{U}_{N, tt}$, $\bar{\xi}^{D}_{N, bb}$, 
$\theta$ in the model III and $\bar{\epsilon}^{U}_{N tt}$, 
$\bar{\epsilon}^{D}_{N bb}$, $\theta$ in the general $3HDM(O_2)$. In our 
numerical calculations, we also respect the constraint for the angle 
$\theta$ due to the experimental upper limit of neutron electric dipole 
moment, $d_n<10^{-25}\hbox{e$\cdot$cm}$, which leads to $\frac{1}{m_t m_b} 
Im(\bar{\xi}^{U}_{N, tt}\, \bar{\xi}^{* D}_{N, bb})< 1.0$ for $M_{H^\pm}
\approx 200$ GeV \cite{david} in the model III and $\frac{1}{m_t m_b} 
(\bar{\epsilon}^{U}_{N,tt}\,\bar{\epsilon}^{D}_{N,bb})\,sin^2\,\theta < 1.0$ 
in $3HDM(O_2)$. Further, we take only $\bar{\xi}^{U}_{N, tt}$ and 
$\bar{\xi}^{D}_{N, bb}$ ($\bar{\epsilon}^{U}_{N, tt}$ and 
$\bar{\epsilon}^{D}_{N, bb}$) nonzero in the model III ($3HDM(O_2)$) and 
neglect all other couplings. Note that $h^0$ is assumed as the lighest Higgs 
boson in our calculations. 

\begin{figure}[htb]
\vskip -3.0truein
\centering
\epsfxsize=6.8in
\leavevmode\epsffile{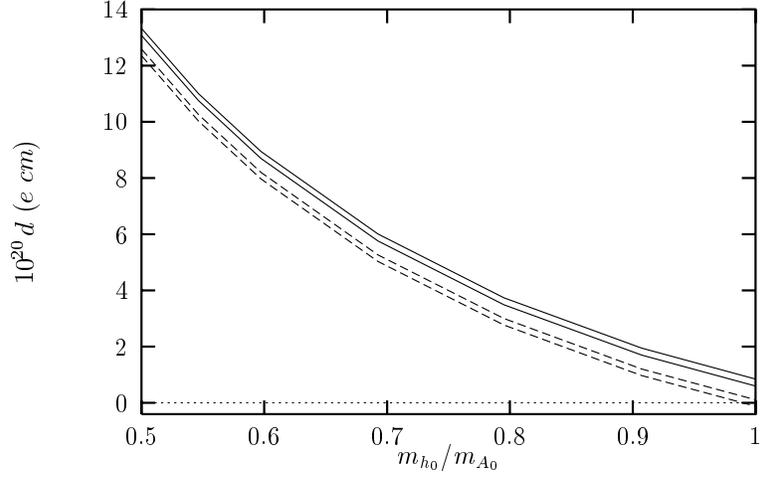}
\vskip -3.0truein
\caption[]{$b$-quark EDM "$d$" as a function of  the ratio $R_{neutr}=
\frac{m_{h^0}}{m_{A^0}}$, for $m_{H^{\pm}}=400\, GeV$, $sin\,\theta = 0.5$, 
$\bar{\xi}_{N,bb}^{D}=40\, m_b$ and $|r_{tb}|=
|\frac{\bar{\xi}_{N,tt}^{U}}{\bar{\xi}_{N,bb}^{D}}| <1$, in the
model III. Here $d$ is restricted in the region bounded by solid lines 
for $C_7^{eff}>0$ and by dashed  lines for $C_7^{eff}<0$.} 
\label{edmr1}
\end{figure}

In  Fig. \ref{edmr1} we plot EDM "$d$" with respect to the ratio 
$R_{neutr}=\frac{m_{h^0}}{m_{A^0}}$ for $sin\,\theta=0.5$, 
$m_{A^0}=80\, GeV$, $m_{H^{\pm}}=400\, GeV$, 
$\bar{\xi}_{N, bb}^{D}=40\, m_b$ and 
$|r_{tb}|=|\frac{\bar{\xi}_{N, tt}^{U}}{\bar{\xi}_{N, bb}^{D}}| <1$ in the
model III. Here "$d$" lies in the region bounded by 
solid lines for $C_7^{eff} > 0$ and by dashed lines for $C_7^{eff} < 0$.
It is observed that $b$-quark EDM is strongly sensitive to the ratio 
$R_{neutr}$ and this dependence increases with decreasing masses of neutral 
Higgs bosons. If the ratio $R_{neutr}$ becomes small the considerable 
enhancement of $d$ will be obtained and therefore the mass difference of 
$h^0$ and $A^0$ should not be large. In the case of degenerate masses of 
$h^0$ and $A^0$, the contribution of the neutral Higgs sector to  $b$-quark 
EDM vanishes.

\begin{figure}[htb]
\vskip -3.0truein
\centering
\epsfxsize=6.8in
\leavevmode\epsffile{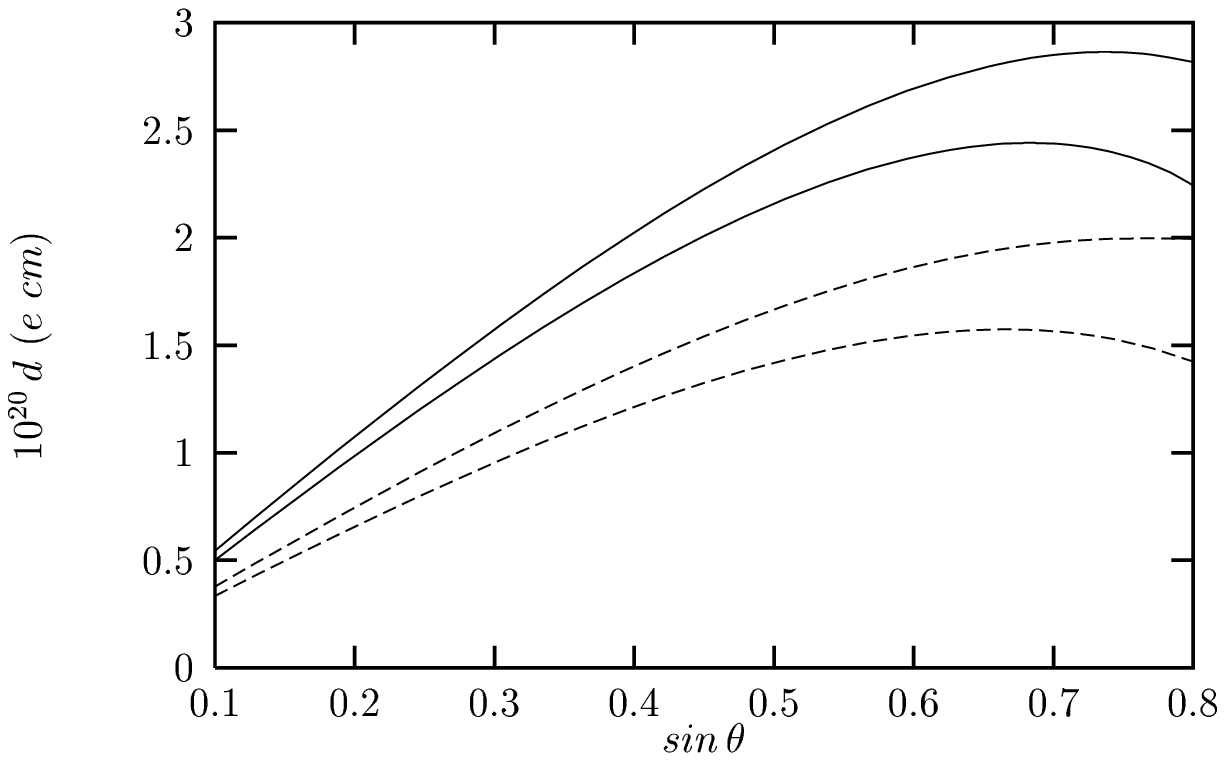}
\vskip -3.0truein
\caption[]{$b$-quark EDM "$d$" as a function of  $sin\,\theta$ 
for $m_H^{\pm}=400\, GeV$, $m_{h^0}=70\, GeV$,  $m_{A^0}=80\, GeV$, 
$\bar{\xi}_{N,bb}^{D}=40\, m_b$ and $|r_{tb}|<1$, in the
model III. Here $d$ is restricted in the region bounded by solid lines 
for $C_7^{eff}>0$ and by dashed  lines for $C_7^{eff}<0$}
\label{edmsin702H}
\end{figure}

\begin{figure}[htb]
\vskip -3.0truein
\centering
\epsfxsize=6.8in
\leavevmode\epsffile{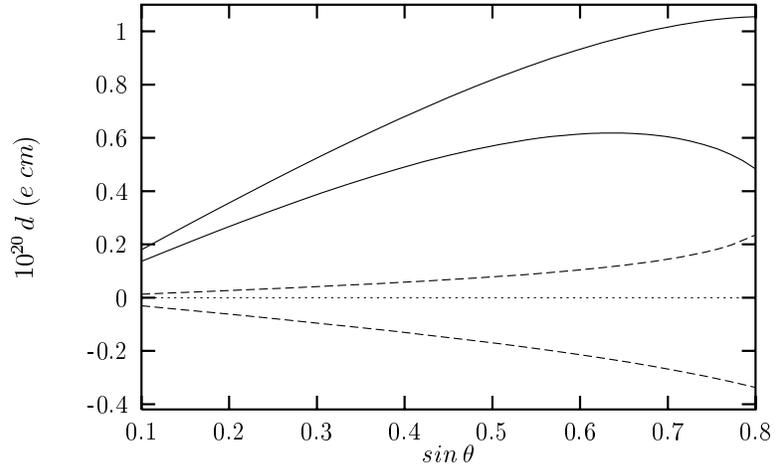}
\vskip -3.0truein
\caption[]{The same as Fig.\ref{edmsin702H} but for $m_{h^0}=m_{A^0}=80\, 
GeV$.}
\label{edmsin802H}
\end{figure}

Fig. \ref{edmsin702H} (\ref{edmsin802H}) is devoted to $sin\,\theta$ 
dependence of "$d$" for $m_{h^0}=70\, GeV$, $m_{A^0}=80\, GeV$ 
($m_{h^0}= m_{A^0}=80\, GeV$), $m_{H^{\pm}}=400\, GeV$, 
$\bar{\xi}_{N, bb}^{D}=40\, m_b$ and $|r_{tb}|<1$. If the value of 
$sin\,\theta$ decreases, "$d$" becomes small as expected and the restricted 
region becomes narrower, for both $C_7^{eff} > 0$ and $C_7^{eff} < 0$. 
"$d$" is positive for $C_7^{eff} > 0$, however, for $C_7^{eff} < 0$, it 
can also take negative values when $m_{h^0}$ reaches to $m_{A^0}$. 
This is an intersting result which can be used in the determination of 
the sign of $C_7^{eff}$.

\begin{figure}[htb]
\vskip -3.0truein
\centering
\epsfxsize=6.8in
\leavevmode\epsffile{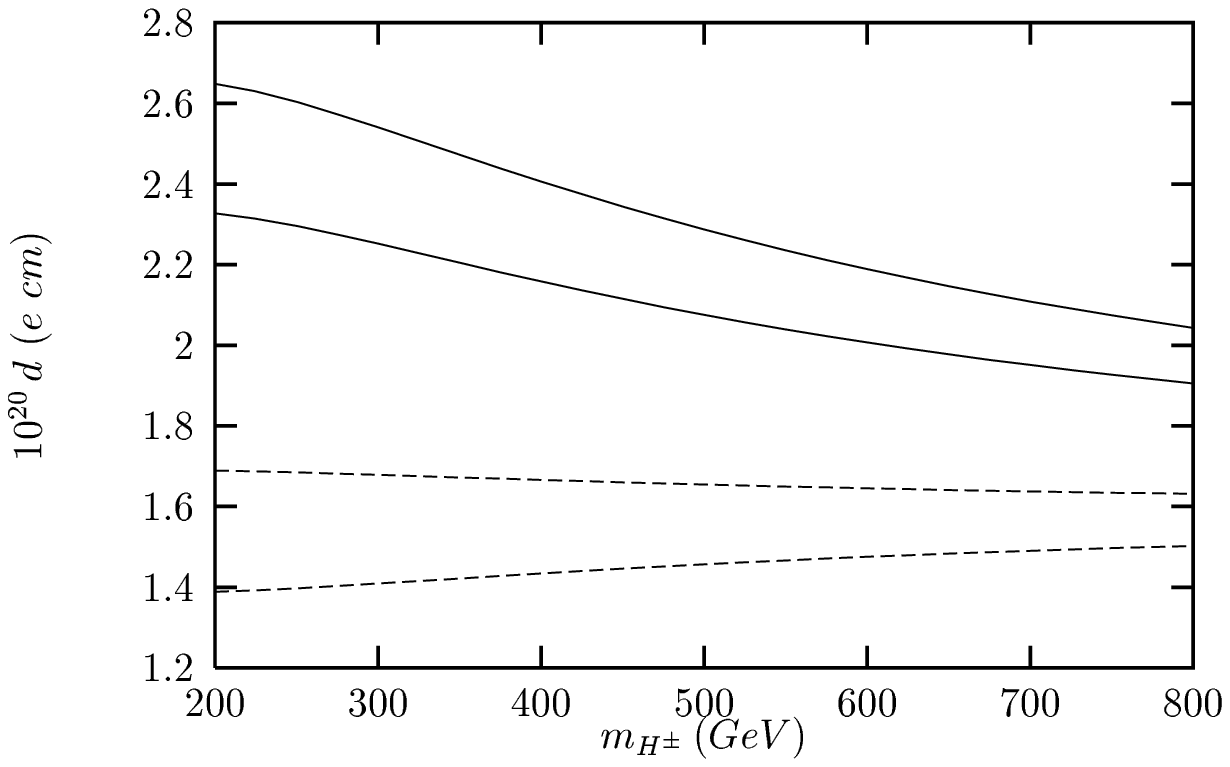}
\vskip -3.0truein
\caption[]{$b$-quark EDM "$d$" as a function of $m_H^{\pm}$, for 
$sin\,\theta = 0.5$, $m_{h^0}=70\, GeV$,  $m_{A^0}=80\, GeV$, 
$\bar{\xi}_{N,bb}^{D}=40\, m_b$ and $|r_{tb}|<1$, in the model III. Here $d$ 
is restricted in the region bounded by solid lines for $C_7^{eff}>0$ and 
by dashed  lines for $C_7^{eff}<0$.}
\label{edm05mh2H}
\end{figure}

In  Fig. \ref{edm05mh2H}, we plot "$d$" with respect to the charged Higgs 
mass $m_{H^{\pm}}$ for $sin\,\theta=0.5$, $m_{h^0}=70\, GeV$, 
$m_{A^0}=80\, GeV$, $\bar{\xi}_{N,bb}^{D}=40\, m_b$ and $|r_{tb}|<1$. 
This figure shows that "$d$" is weakly sensitive to the charged Higgs mass 
$m_{H^{\pm}}$ for $m_{H^{\pm}}\geq 400\,GeV$, especially in the case 
$C_7^{eff} < 0$. The restriction region becomes narrower with increasing 
$m_{H^{\pm}}$.

In $3HDM (O_2)$, the possible choice of the parametrization for Yukawa 
couplings (eq. (\ref{yukpar})) causes to vanish the contribution of the 
neutral Higgs sector to the $b$-quark EDM (see eq. (\ref{EDM3H3})) 
and therefore only the charged Higgs sector contributes. 

\begin{figure}[htb]
\vskip -3.0truein
\centering
\epsfxsize=6.8in
\leavevmode\epsffile{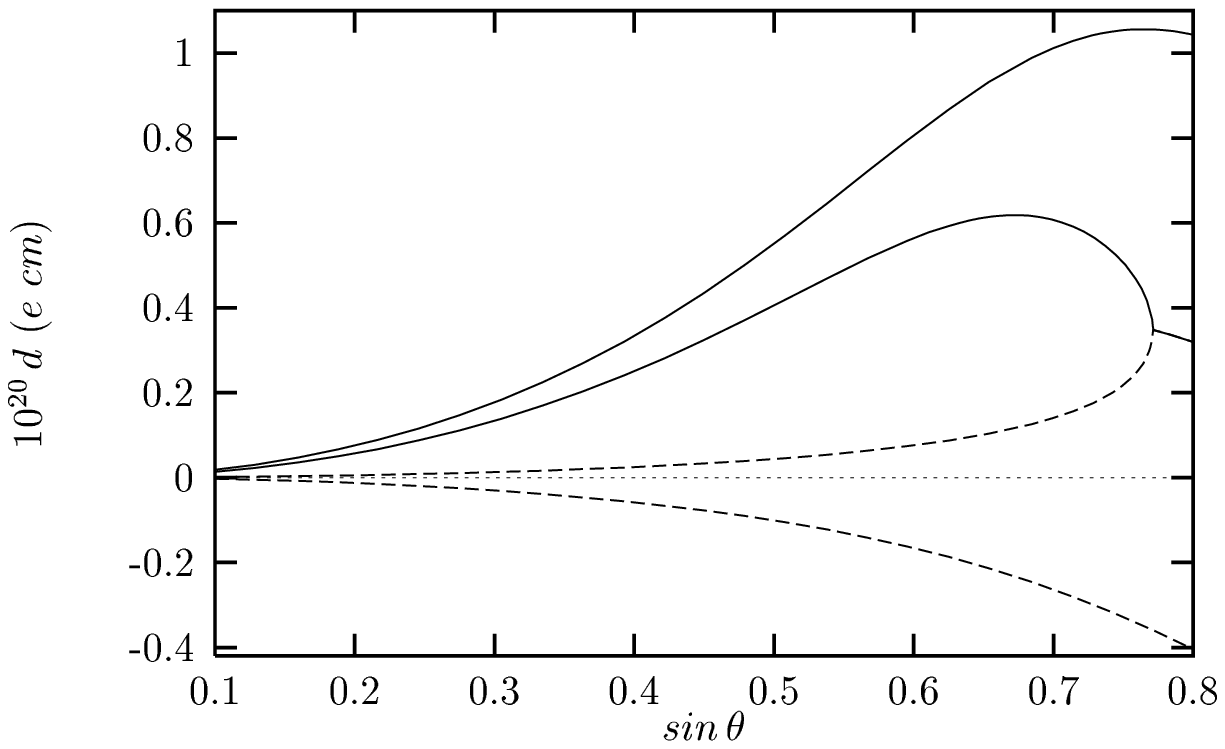}
\vskip -3.0truein
\caption[]{The same as Fig. \ref{edmsin802H}, but in $3HDM(O_2)$.}
\label{edmsin3H}
\end{figure}

\begin{figure}[htb]
\vskip -3.0truein
\centering
\epsfxsize=6.8in
\leavevmode\epsffile{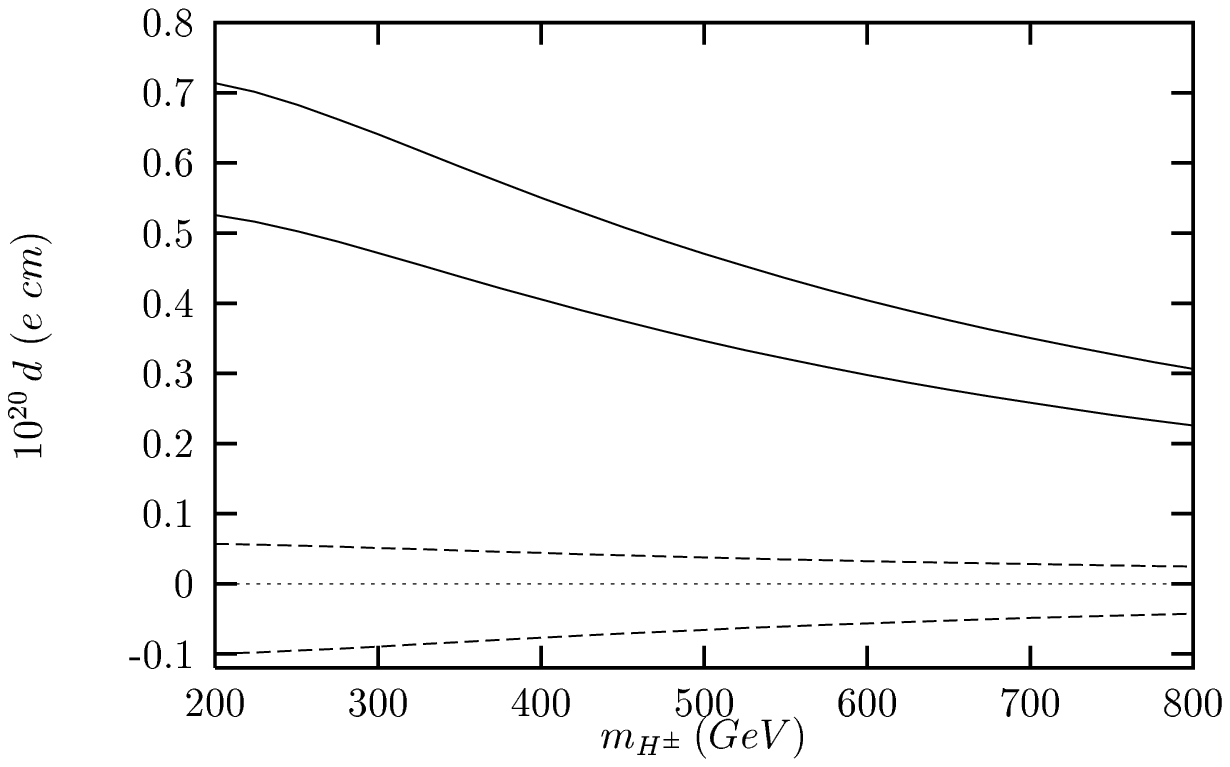}
\vskip -3.0truein
\caption[]{The same as Fig. \ref{edm05mh2H}, but in $3HDM(O_2)$.}
\label{edm05mh3H}
\end{figure}

Fig. \ref{edmsin3H} is devoted to $sin\,\theta$ dependence of "$d$" for 
$m_{H^{\pm}}=400\, GeV$, $\bar{\xi}_{N,bb}^{D}=40\, m_b$ and $|r_{tb}|<1$
in $3HDM (O_2)$. The behaviour of EDM is similar to the model III case with 
degenerate masses $m_{h^0}$ and $m_{A^0}$, however, "$d$" is more sensitive 
to $sin\,\theta$, since it is proportional to $sin^2\,\theta$ 
(see eq. (\ref{EDM3H3})). Further, "$d$" is not sensitive to the charged 
Higgs mass $m_{H^{\pm}}$ for $m_{H^{\pm}}\geq 400\,GeV$, 
especially in the case $C_7^{eff} < 0$, similar to the model III. 
(Fig. \ref{edm05mh3H})

Now we would like to summarize the main points of our results:

\begin{itemize}

\item It is interesting that EDM is generated by the one loop diagrams. This
is due to the freedom to choose the Yukawa couplings as complex numbers in 
the models under consideration. Further, even in the model with two Higgs
doublets, there exist a non-vanishing contribution to EDM due to the charged 
sector.

\item Since $b$-quark EDM "$d$" is strongly sensitive to the ratio 
$R_{neutr}$, in model III, the mass difference of $h^0$ and $A^0$ should not
be large. For the 3HDM under consideration, there is no need to restrict the 
neutral Higgs  masses because they do not contribute to "$d$" for the given 
parametrization of Yukawa couplings.    

\item  If "$d$" is positive, $C_{7}^{eff}$ can have both signs. However, if 
it is negative, $C_{7}^{eff}$ must be negative for both models. This is an 
important observation which is useful in the determination of the sign of 
$C_{7}^{eff}$.

\item "$d$" is not sensitive to the mass of charged Higgs boson for its
large values in both models. 

\item EDM of "$b$" quark is at the order of $\sim 10^{-20} e \, cm$ 
in both model III and $3HDM(O_2)$ and its magnitude is larger compared to
the results ($\sim 10^{-23}-10^{-22} e \, cm$) existing in the literature. 
The neutral Higgs boson effects are strong in the model III and there is 
an enhancement, even one order ($d\sim(10^{-19}$), if the masses of neutral 
Higgs bosons, $m_{h^0}$ and $m_{A^0}$, are far from degeneracy. 

\end{itemize}

Therefore, the experimental investigation of the $b$-quark EDM gives 
powerful informations about the physics beyond the SM.

\newpage
\newpage

\newpage
\end{document}